\documentstyle[11pt]{article}

\begin{document}

\setlength{\baselineskip}{1.0cm}

\begin{center}
{\Huge \bf Axial Monopoles, Quantization of Electric Charge and Dynamical
Discreteness of Space-Time}
\end{center}

\vspace{1.0cm}
{\LARGE
\begin{center}
{\bf M.C. Nemes$^{(1)}$ and
Saulo C.S. Silva$^{(2)}$}
\end{center}

\begin{center}
$^{(1)}$Instituto de Ci\^encias Exatas, Universidade Federal de Minas
Gerais\\
\end{center}

\begin{center}
$^{(2)}$Instituto de F\'{\i}sica, Universidade de S\~ao Paulo\\
\end{center}
\vspace{0.5cm}

\begin{abstract}
In the present contribution we show that the introduction of axial currents
in electrodynamics can explain the quantization of electric charge and
introduces a dynamical discreteness of space-time, justifying thus the
regularization of Feymman's integrals.
\end{abstract}
\vspace{0.5cm}

We start with a generalized definition of the electromagnetic field tensor

\begin{equation}
F_{\mu\nu}=\partial_{\mu}A_{\nu}-\partial_{\nu}A_{\mu}
+\epsilon_{\mu\nu\alpha\beta}\partial^{\alpha}B^{\beta}
\end{equation}

\vspace{1.0cm}

\noindent where $B^\mu$ is an axial 4-vector field. From the point
of view of quantum theory this field represents photon-like particles
except for $P$, $T$ and $C$ parities. In other words, axial photons.

Maxwell's equations in Lorenz's gauge become

\begin{eqnarray}
\partial^{\nu}F_{\nu\mu}=\Box A_{\mu}=j_{\mu}\\
\partial^{\nu}F_{\nu\mu}^{\dagger}=\Box B_{\mu}=g_{\mu}
\end{eqnarray}

\vspace{1.0cm}

\noindent where we have introduced the axial electromagnetic current given by

\begin{equation}
g_{\mu}=-g\bar{\psi}\gamma_{\mu}\gamma_{5}\psi
\end{equation}

\vspace{1.0cm}

\noindent Here $\psi$ represents an spin $1/2$ particle
(axial monopole) with
axial charge $g$.

In terms of the electromagnetic fields, equation $(3)$ is written as

\begin{eqnarray}
\vec{\nabla}\cdot\vec{H} & = & -g\bar{\psi}\gamma_{0}\gamma_{5}\psi=
-g\psi^{\dagger}\gamma_{5}\psi\\
\vec{\nabla}\times\vec{E} & = & -\frac{\partial\vec{H}}{\partial t}
-g\bar{\psi}\vec{\gamma}\gamma_{5}\psi=
-\frac{\partial\vec{H}}{\partial t}
-g\psi^{\dagger}\vec{\alpha}\gamma_{5}\psi
\end{eqnarray}

\vspace{1.0cm}

Let us consider an axial monopole at rest. Equation $(6)$ gives

\begin{eqnarray}
\vec{\nabla}\times\vec{E}=\rho\vec{\sigma}
\end{eqnarray}

\vspace{1.0cm}

\noindent where $\rho$ stands for the axial charge density and $\vec{\sigma}$
is the expectation value of the spin operator. This equation has the solution

\begin{equation}
\vec{E}= \frac{g}{4\pi}\vec{\sigma}\times \frac{\vec{r}}{r^3}
\end{equation}

\vspace{1.0cm}

Consider the scattering of an electric charge by this field, at impact
parameter $b$, with $\vec{\sigma}$
pointing in the direction of the charge's velocity $\vec{v}$
(positive z-axis, say). In such case,
the variation of the charge's momentum during the scattering can be immediately
calculated to be

\begin{equation}
\Delta\vec{p}=\int_{-\infty}^{+\infty}\vec{F}dt=\int_{-\infty}^{+\infty}
e\vec{E}dt=\int_{-\infty}^{+\infty}
\frac{eg}{4\pi}\frac{\rho}{(\rho^{2}+z^{2})^{\frac{3}{2}}}\hat{e}_{\phi}dt
\end{equation}

\vspace{1.0cm}

Inserting $\rho=b$ and $z=vt$ in $(9)$, we get

\begin{equation}
\Delta\vec{p}=\frac{egb}{4\pi}\hat{e}_{\phi}\int_{-\infty}^{+\infty}
\frac{dt}{(b^{2}+v^{2}t^{2})^{\frac{3}{2}}}=
\frac{eg}{2\pi vb}\hat{e}_{\phi}
\end{equation}

\vspace{1.0cm}

To this momentum variation there will be a corresponding angular momentum's
change given by

\begin{equation}
\Delta\vec{L}=\frac{eg}{2\pi v}\hat{e}_{z}
\end{equation}

\vspace{1.0cm}

\noindent which is independent of the impact parameter.

Using now Bohr's quantization rule, we arrive at

\begin{equation}
\Delta L = \frac{eg}{2\pi v} = n
\end{equation}

\vspace{1.0cm}

\noindent $n$ being any integer.

The above relation can only be satisfied if we simultaneously fulfill

\begin{equation}
\frac{eg}{2\pi}=n_0
\end{equation}

\noindent and

\begin{equation}
v=\frac{n_0}{n}
\end{equation}

\vspace{1.0cm}

\noindent with $n = n_0$, $n_0+1$, $n_0+2$...

\vspace{1.0cm}

The first one of these conditions implies in charge quantization, in the
same way of Dirac's charge quantization condition.

The second condition
represents the restriction of velocity values to rational numbers, a result
integrated in discrete space-time theories. Moreover,
for a massive charged particle, it means that $v\leq v_0$

\begin{equation}
v_0=\frac{n_0}{n_0+1}
\end{equation}

\vspace{1.0cm}

This limit for $v$ leads to upper limits for $p$ and $E$, the momentum and
energy of such particle. For $n_0\gg 1$ these upper limits are

\begin{equation}
p_0\approx E_0\approx m(n_0/2)^{\frac{1}{2}}
\end{equation}

\vspace{1.0cm}

Using the uncertainty principle, we arrive at a fundamental length for the
particle's space-time, namely

\begin{equation}
a \sim \frac{1}{p_0} \approx \frac{(2/n_0)^\frac{1}{2}}{m}
\end{equation}

\vspace{1.0cm}

The experimental upper limit $a < 10^{-16} cm$ gives

\begin{equation}
n_0 > 2 \times 10^6
\end{equation}

\noindent and

\begin{equation}
g=\frac{2\pi n}{e} > 10^9
\end{equation}

\vspace{1.0cm}

\noindent where was used the mass and charge of the electron.

With this value we can estimate a lower limit to the mass of the axial monopole

\begin{equation}
M=\frac{g^2}{e^2} m > 10^{16} Gev
\end{equation}

\vspace{1.0cm}

This limit is of order of GUT scale, the order of, for instance,
SU(5) magnetic monopoles.

The dynamical discreteness obtained here can be used to justify the
regularization of the Feymman's integrals of QED, introducing
the covariant cut-off
given by $(16)$.

For example, we can obtain the
relation between the physical and bare values of the fine structure constant

\begin{equation}
\alpha = \alpha_0 [1-\frac{\alpha_0}{3\pi}
log (\frac{m^2}{p_0^2})]^{-1}= \alpha_0 [1-\frac{\alpha_0}{3\pi}
log (2/n_0)]^{-1}
\end{equation}

\vspace{1.0cm}

If we take the lower limit for $n_0$ given by $(18)$, we get

\begin{equation}
\alpha = \alpha_0 [1+\frac{6\alpha_0}{3\pi} log 10]^{-1} \approx
\frac{\alpha_0}{1+ 1.5 \alpha_0}
\end{equation}

\vspace{1.0cm}

Thus, we conclude that the existence of massive axial monopoles in
the Universe allows us to explain the quantization of electric charge
and gives rise to a dynamical, covariant, discreteness of space-time.
}

\end{document}